\documentstyle[twocolumn,prl,aps,epsf]{revtex}

\begin{document}

\twocolumn[
\hsize\textwidth\columnwidth\hsize\csname@twocolumnfalse\endcsname
\draft
\title{Electron Spin Resonance of defects in the Haldane System Y$_2$BaNiO$_5$}
\author{C.D. Batista,  K. Hallberg and A.A.Aligia}
\address{Centro At\'omico Bariloche and Instituto Balseiro}
\address{Comisi\'on Nacional de Energ{\'\i}a At\'omica}
\address{8400 S.C. de Bariloche, Argentina.}
\date{Received \today }
\maketitle

\begin{abstract}
We calculate the electron paramagnetic resonance (EPR) spectra of the antiferromagnetic spin-1
chain compound Y$_{2}$BaNi$_{1-x}$Mg$_{x}$O$_{5}$ for different values of $x $
and  temperature $T$ much lower than the Haldane gap ($\sim$ 100K). The low-energy spectrum
 of an anisotropic Heisenberg Hamiltonian, with all parameters determined from
 experiment, has been solved using DMRG.
The observed EPR spectra  are quantitatively reproduced by this model.
The presence of end-chain $S=1/2$ states is clearly observed as the main peak in
 the spectrum and the remaining structure is completely understood.
\end{abstract}


]

\narrowtext

A great deal of interest in one dimensional (1D) Heisenberg chains with
nearest-neighbor antiferromagnetic exchange coupling, $J$, has been
originated by Haldane's conjecture that integer-valued spin chains would
exhibit a gap in the spin-wave excitation spectrum \cite{hal}. An
interesting model of the underlying wave function for the Haldane state has
been proposed by Affleck {\it et al. }\cite{aff}. They have shown that the
exact ground state of the Hamiltonian $\sum_{i{\bf S}_{i}\cdot {\bf S}i+1}+(%
{\bf S}_{i}\cdot {\bf S}_{i+1})^{2}/3$ is a valence-bond-solid (VBS) state.
This state in an open chain has two unpaired $S=1/2$ spins, one at each end.
In agreement with this, exact diagonalization \cite{ken} of finite open
Heisenberg chains (without the biquadratic term) shows that the four
lowest-lying states are a triplet and singlet whose energy separation
approaches zero exponentially with increasing length. Monte Carlo \cite{miy}
and density-matrix renormalization-group (DMRG) \cite{whi} studies clearly
show the presence of $S=1/2$ end states.

This picture was supported by EPR measurements of [Ni(C$_{2}$H$_{8}$N$_{2}$)$%
_{2}$(NO$_{2}$)]ClO$_{4}$ (NENP) doped with non-magnetic ions \cite{gla},
where resonances corresponding to the fractional spin $S=1/2$ states at the
``open'' ends of the $S=1$ Ni chains were observed. Similar measurements for
doping with magnetic ions are also consistent with $S=1/2$ end states \cite
{aff2}. This effect, if robust, would correspond to the only instance in
magnetism where a low energy collective excitation has no classical analog.

However, Ramirez {\it et al.} \cite{ram} also tested the presence of free $%
S=1/2$ states by studying the specific heat of non-magnetic defects in Y$%
_{2} $BaNiO$_{5}$, with magnetic fields up to 9T and temperatures down to
0.2K. They found that the shape and magnitude of the Schottky anomaly
associated with the defects in Y$_{2}$BaNi$_{1-x}$Zn$_{x}$O$_{5}$ are better
described by a simple model involving spin-1 excitations, instead of the $%
S=1/2$ excitations of the VBS. In order to eliminate the apparent
discrepancy with the EPR measurements in NENP, Ramirez {\it et al.} \cite
{ram} pointed out the possibility that a small fraction of ethylene diamine
complexes in NENP acquire charge at structural defects induced by Zn doping.
As it is not uncommon that structural defects induce a paramagnetic behavior
in organic compounds, they propose that similar EPR measurements should be
performed on Y$_{2}$BaNi$_{1-x}$A$_{x}$O$_{5}$ (A a nonmagmetic ion) where $%
x $ is more easily calibrated \cite{ram}. EPR is an appropriate technique to
determine the existence of S=1/2 end states because it allows to distinguish
between S=1/2 and S=1 spins in the presence of spatial anisotropy.

In a recent paper \cite{nos1}, a precise fit of the above mentioned specific
heat measurements was done, solving the low-energy spectrum of an
anisotropic Heisenberg Hamiltonian for Y$_{2}$BaNi$_{1-x}$Zn$_{x}$O$_{5}$.
It was shown that there is no contradiction between EPR in NENP \cite{gla}\
and specific heat measurements in Y$_{2}$BaNi$_{1-x}$Zn$_{x}$O$_{5}$ \cite
{ram}. The results supported the existence of $S=1/2$ excitations for
sufficiently long chains and clearly indicated that the anisotropy plays a
very important role in the low temperature properties. However, EPR
experiments would unambiguously detect the presence of such excitations and
definitely confirm the validity of the theory.

Recently, Saylor {\it et al }\cite{say} have measured the EPR spectra of Y$%
_{2}$BaNi$_{1-x}$Mg$_{x}$O$_{5}$ at temperature $T=2K$ for different values
of Mg concentration. The spectra show a prominent main peak, which can be
associated with free S=1/2 spins, surrounded by some secondary peaks. The
appearance of the latter is not related with the interaction between the
S=1/2 end states belonging to different chains. Thus, it is necessary to
give an explanation for the origin of these other resonances.

In this paper, we calculate the EPR spectra of Y$_{2}$BaNi$_{1-x}$Mg$_{x}$O$%
_{5}$ at $T$=2K for different values of $x$ ($x=0.002,\;0.006,\;$and $0.02$)
and compare them with experiment. Mg and Zn are non-magnetic impurities, so
the substitution of these ions for Ni provides a simple break in the chain.
Due to the high sensitivity of EPR experiments, it is optimal for discerning
between long chains (where the $S=1/2$ end spins are nearly independent) and
the shorter chains (where the $S=1/2$ objects are overlapping). We use the
same Hamiltonian as in Ref. \cite{nos1}, with the same parameters as those
taken there from inelastic neutron scattering \cite{sak,xu}. The calculated
EPR spectra show several contributions from chains with different lengths
which are quantitatively reproduced. These contributions give rise to the
secondary peaks mentioned above. The location of these peaks are related to
the effect of anisotropy on chains with different lengths. In this way,
these secondary peaks constitute clear evidence that the observed spectrum
is originated by the end states of the 1D NiO$_{5}$ chains and not by
paramagnetic impurities.

Y$_{2}$BaNiO$_{5}$ has an orthorhombic crystal structure with the Ni$^{2+}$ (%
$S=$1) ions arranged in linear chains with a nearest-neighbor
antiferromagnetic superexchange coupling $J$. The interchain coupling $%
J_{\perp }$ is at least three orders of magnitude weaker, making this
compound an ideal one-dimensional antiferromagnetic chain. While each Ni
atom is surrounded by six O atoms in near octahedron coordination, the true
site symmetry is $D_{2h}$. Neglecting $J_{\perp }$ (the effect of which will
be discussed later), the appropriate Hamiltonian for a chain segment of
length $N$ between two Mg impurities is \cite{sak,xu,gol}:

\begin{eqnarray}
H &=&\sum_{i}\{J{\bf S}_{i}\cdot {\bf S}%
_{i+1}+D(S_{i}^{z})^{2}+E[(S_{i}^{x})^{2}-(S_{i}^{y})^{2}]\}  \nonumber \\
&&-\mu _{B}\sum_{\upsilon \alpha }B^{\alpha }g^{\alpha \nu }S_{t}^{\nu }
\label{ecu1}
\end{eqnarray}
where $z$ is along the chain axis, ${\bf S}_{t}$ is the total spin and $%
g^{\alpha \nu }$ is the gyromagnetic tensor . Recent estimates based on fits
of the Haldane gaps (in $x,\;y,\;$and $z$ directions) measured by inelastic
neutron scattering , indicate $J\sim 280K\;$, $D\sim -.038J\;$, and $E\sim
-.0127J\;$\cite{sak,xu}. As Y$_{2}$BaNiO$_{5}$ has a Haldane gap $\sim 100K$ 
\cite{sak,xu}, the spin wave contribution to the EPR is negligible at 2$K$.
In this temperature range, EPR is dominated by the effect of the defects.
For this reason, it is necessary to solve accurately the low energy spectra (%
$\omega \leq 10K$) of $H$ for all values of $N$ to be able to explain the
observed EPR\ experiment.

The lowest energy states of Eq.(\ref{ecu1}) are a spin singlet, which we
denote as $|0\rangle $, and three components of a split spin triplet ($%
|1S_{t}^{z}\rangle $). The energy of $|11\rangle $ is the same as that of $%
|1-1\rangle $ due to the time-reversal symmetry of $H$. By means of the DMRG
method, we have calculated the energy of these four states for all $N\leq 40$
and $E=0.$ The difference of energy $e$ between any pair of eigenstates
decays exponentially to zero with increasing $N$. This behavior allows us to
extrapolate the energies to all values of $N>40$, and demonstrate that the $%
S=$1/2 spins at the end of the chains are also asymptotically free in the
presence of anisotropy. This issue is easy to understand considering that $%
S= $1/2 spins cannot be affected by anisotropy due to time-reversal
symmetry. At this level, it is important to remark that the two end $S=$1/2
spin excitations, which are localized at the end sites in the VBS, have a
finite localization length $l\sim $ 6 sites for the pure Heisenberg model 
\cite{ken,miy,whi}. Therefore, while they are nearly free for large ($N>>l$)
open chains, the interaction between them is considerable when the length of
the chain $N$ is comparable to $2l.$ This interaction splits the two $S=$1/2
states into a singlet (ground state for even $N$) and a triplet (ground
state for odd $N$) \cite{lie}.

The difference between any two energies of the above mentioned low-energy
states is linear in $D$, and the quadratic corrections are negligible \cite
{gol}. This widely justifies the validity of perturbation theory to first
order in $D$. Then, by symmetry we can also include the term $%
\sum_{i}E[(S_{i}^{x})^{2}-(S_{i}^{y})^{2}]$ to first order. Thus we find the
following low-energy effective Hamiltonian including the triplet $\left|
1S_{z}\right\rangle $ and the singlet state $\left| 0\right\rangle $:

\begin{eqnarray}
H_{eff} &=&E_{0}(N)+(J\alpha (N)+D\beta (N))\left| 0\right\rangle
\left\langle 0\right| +D\gamma (N)S_{z}^{2}  \nonumber \\
&&+E\gamma (N)(S_{x}^{2}-S_{y}^{2})-\mu _{B}\sum_{\nu \alpha }B^{\alpha
}g^{\alpha \nu }S_{t}^{\nu }  \label{ecu2}
\end{eqnarray}
where $E_{0}(N),\;\alpha (N)$, $\beta (N)$ and $\gamma (N)$ are functions of
the chain length $N$, determined from the DMRG data fitting exactly the four
lowest energy levels for $E=0$. The numbers $\alpha (N)$, $\beta (N)\;$and $%
\gamma (N)$ are represented in Fig. 1. The validity of the last term of Eq.(%
\ref{ecu2}) has been verified explicitly by calculating the matrix elements
of $S_{t}^{+}$, and $S_{t}^{-}$ for all chains. $H_{eff}$ determines the
thermodynamics of the system at temperatures well below the Haldane gap.

\begin{figure}[htbp]
\vspace*{-0.4cm}
\epsfxsize=3.3in
\epsfysize=2.75in
\epsffile{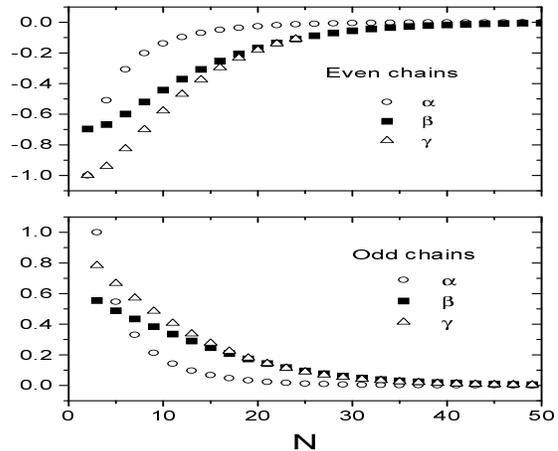}
\vspace*{-0.1cm}
\caption[]{{ Numerical coefficients of the effective Hamiltonian
Eq. (\ref{ecu2}%
) as functions of the chain length. }}
\end{figure}

Assuming a random distribution of defects, the probability to find a Mg atom
at any site $i$ followed by $N$ Ni sites and another Mg atom at site $i+N+1$
is clearly $x^{2}(1-x)^{N}$ , where $x$ is the concentration of Mg (missing $%
S=1$ Ni spins). Thus, the probability per Mg impurity of finding a segment
of $N$ Ni atoms is $x(1-x)^{N}.$ Then, the EPR signal intensity $I(x,\omega
,T)$ per impurity is:

\begin{equation}
I(x,\omega ,T)=\sum_{N=1}^{\infty }\;x(1-x)^{N}\;I_{N}(\omega ,T)
\label{ecu3}
\end{equation}
where $I_{N}(\omega ,T)$ is the EPR signal intensity of a segment of length $%
N$ described by $H_{eff}$. This intensity is given by the following
expression:

\begin{equation}
I_{N}(\omega ,T)\propto \sum_{\alpha ,\nu =1}^{4}\left| \left\langle \nu
\left| {\bf S\cdot n}\right| \alpha \right\rangle \right| ^{2}e^{-\beta
E_{\alpha }}\;\delta (\omega -E_{\nu }+E_{\alpha })  \label{ecu4}
\end{equation}
where $\left| \alpha \right\rangle $ are the eigenstates of $H$ (or $H_{eff}$%
) for the $N$-site chain, {\bf n }is the direction (orthogonal to {\bf B) }%
of the{\bf \ }applied microwave magnetic field and $\beta$ the inverse
temperature. The delta distributions have been replaced by Lorentzian
functions with a finite width to simulate the effect of interactions not
included in Eq.(\ref{ecu2}) and the experimental resolution.

\begin{figure}[htbp]
\vspace*{-0.4cm}
\epsfxsize=3.3in
\epsfysize=2.75in
\epsffile{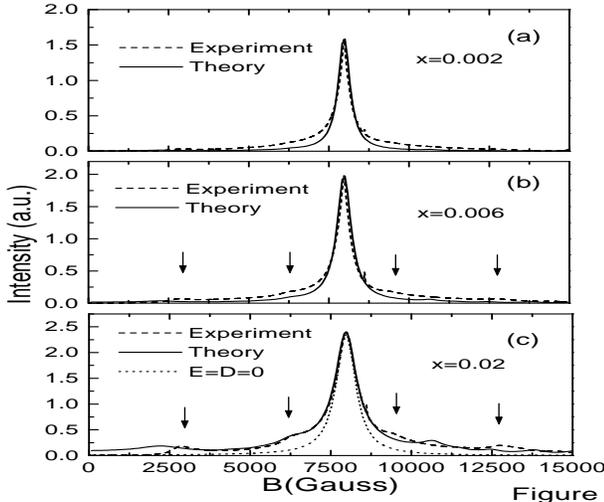}
\vspace*{-0.1cm}
\caption[]{{Comparison between the experimental and theoretical
EPR spectra
for three different concentrations of Mg, in the absence of interchain
coupling. The values of $g^{aa}=2.254,\;g^{bb}=2.177,\;$and $g^{cc}=2.165$
are taken from Ref. \cite{say}{\bf .} The parameters of the Hamiltonian 
(Eq. \ref{ecu1}) are taken from neutron scattering experiments
\cite{sak,xu}%
: $J=280K,\;D=-0.039J,\;E=-0.127J$. The arrows indicate the positions of
the
secondary peaks. }}
\end{figure}

In Fig. 2 we show the experimental data \cite{say} for $I(x,T)$ with three
different values of $x\;(x=0.002,\;0.006,\;$and$\;0.02)$ at $T$=2K and
compare them with the calculations. The uniform magnetic field ${\bf B}$
points in the {\bf x (}or{\bf \ c)} direction, while the direction of the
microwave field is {\bf n=z} (or {\bf a}). The width and intensity of the
lorentzians corresponding to each transition have been adjusted considering
the main peak. These are the only adjustable parameters. The integrated
intensity is determined by $x.$ The most prominent feature of Fig. 2 is the
main peak at around 8000G which is related to very long chains for which the
splitting induced by the anisotropy terms is much smaller than $g^{cc}\mu
_{B}B$. Surrounding the main peak there are some secondary peaks which
correspond to the chains for which the magnitude of the anisotropy splitting
is of the same order as $g^{cc}\mu _{B}B.$ In order to clarify the last
statement, let us consider the simplest case of $E=0$. The splitting $\Delta
(N)$ (between the $S^{z}=0$ and the $S^{z}=\pm 1$ states) induced by the
anisotropy term $DS_{z}^{2}$ ($D<0$) for odd and even chains is represented
schematically in Fig. 3. With respect to these latter states, a uniform
magnetic field ${\bf B}$ along the ${\bf x}$ direction (perpendicular to the
chains), does not shift the $S_{x}=0$ state ($(\left| 11\right\rangle
-\left| 1-1\right\rangle )/\sqrt{2}$) but mixes $\left| 10\right\rangle $
and $\frac{1}{\sqrt{2}}(\left| 11\right\rangle +\left| 1-1\right\rangle $
giving rise to a splitting between the lowest triplet states T$_{1}$ and $%
S_{x}=0$: $\delta (N)=\Delta (N)/2+\sqrt{(\Delta (N)/2)^{2}+(g^{cc}\mu
_{B}B)^{2}}.$ The application of a microwave magnetic field along the
direction of the chains will induce transitions between these two states
giving rise to resonant peaks when $\hbar \omega =\delta (N),$ being $\omega 
$ the microwave frequency. Then, chains larger than 40 sites ($\Delta
(N)<<g^{cc}\mu _{B}B$) contribute to the main peak centered at $\hbar \omega
=g^{cc}\mu _{B}B$. For chain lengths between 15 and 30, $\Delta (N)$ is of
the same order as $g^{cc}\mu _{B}B$ giving rise to some secondary peaks at
higher and lower energies than $g^{cc}\mu _{B}B$. $\Delta (N)$ is negative
for odd chains and positive for even ones, so the peaks on the left side
with respect to the main peak come from even chains while the ones on the
right side are produced by odd chains. In Fig. 2 c) we have also plotted the
absorption curve corresponding to no anisotropy ($D=E=0$). As is expected
from the analysis given above, no feature is observed apart from the main
peak. We also note that the weight of the secondary peaks is bigger for
larger doping concentration ($x=0.02$). This is due to the fact that at
larger doping, the probability of finding shorter chains increases (see Eq.
3). The discrepancy between theory and experiment at smaller $x$
(particularly 0.002) might be due to uncertainties in the actual
concentration of impurities \cite{say}.

\begin{figure}[htbp]
\vspace*{-0.4cm}
\epsfxsize=3.3in
\epsfysize=2.75in
\epsffile{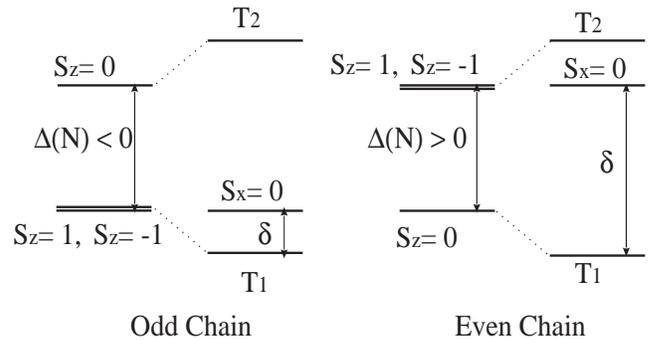}
\vspace*{-1.cm}
\caption[]{{Diagram which shows the effect on the triplet states of an
applied
magnetic field ${\bf B}$ in the $x$ direction in the presence of the
anisotropy term $S_{z}^{2}.$ }}
\end{figure}

The calculated spectra are very sensitive to the values of $D$ and $E.$
However the best agreement is obtained using the experimental values given
above \cite{sak,xu}. In order to show the sensitiveness of the spectra to
these parameters, in Fig.4 we plot the spectrum for $x=0.02$ and two
different sets of parameters.

\begin{figure}[htbp]
\vspace*{-0.4cm}
\epsfxsize=3.3in
\epsfysize=2.75in
\epsffile{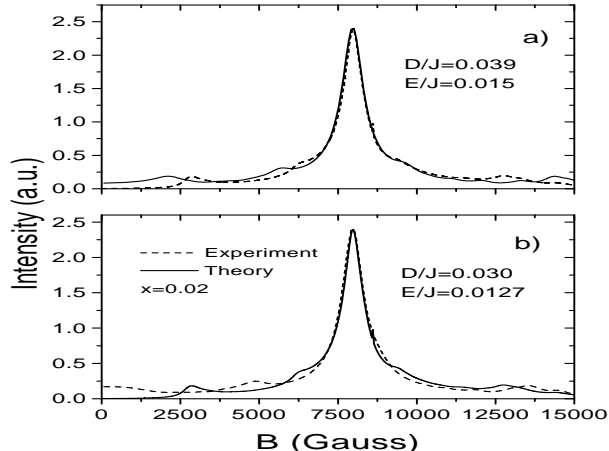}
\vspace*{-0.1cm}
\caption[]{{same as Fig.2 for $x=0.02$ and a) $D=0.039J$, $E=0.015J$; b) $%
D=0.030J$, $E=0.0127J.$ }}
\end{figure}

The position of the different peaks is well reproduced within an error of
1000G$\sim 1K$ (the spike at around 8500G is a calibration feature). This
error is of the same order of magnitude as the interactions neglected in our
theory \cite{note1}. It is important to note that there are no fitting
parameters changing the positions of the secondary peaks. This agreement
shows clearly that the secondary peaks of the EPR spectra are originated by
the effect of spatial anisotropy on chains with lengths between 15 and 40
sites. Due to experimental difficulties at very low fields, there is an
artificial cutoff below 2500G.

In Ref. \cite{nos1}, we have included the interactions between chain
segments in order to reproduce the specific heat measurements of Ref. \cite
{ram} at zero magnetic field. Because of this it is important to determine
how sensitive the calculated EPR spectra are to the inclusion of these
interactions. The origin and the magnitude of this $J^{\prime }$ inter-chain
interaction is explained in Refs. \cite{nos1,bat}. Since a calculation of
all possible $J^{\prime }$s and their effect on a 2D (or 3D) topology is a
formidable task, we model the effect of $J^{\prime }$ considering a
collection of chain segments with effective 1D topology with exchange
interaction $J^{\prime }$, coming from a uniform distribution between $%
-J_{\max }^{\prime }$ and $J_{\max }^{\prime }.$ We obtain that the
inclusion of realistic values of $J^{\prime }$ has no effect on the position
of the peaks. The only effect of $J^{\prime }$, as might be expected, is a
broadening of the secondary peaks and the corresponding small reduction of
their maxima. As a consequence, the agreement between theory and experiment
slightly improves for the peaks near 6000G and 12500G.

In conclusion, by solving the low-energy spectrum of a Heisenberg
Hamiltonian $H$ which includes experimental axial and planar anisotropy, we
have reproduced in detail the low-temperature EPR data measured in Y$_{2}$%
BaNi$_{1-x}$Mn$_{x}$O$_{5}$ for different concentrations of Mg$.$ It is
important to remark that {\em there are no fitting parameters to reproduce
the positions of the different peaks in the EPR spectra}. These results
confirm that the $S=$1/2 end chain excitations, which are asymptotically
free for large chain segments, are experimentally observed. The fact that we
have been able to fit all peaks with this model indicates that the $S=1/2$
feature observed by EPR experiments is unambiguously an end-chain
excitation, ruling out any other possibility. Summarizing, the EPR
experiments can be interpreted in terms of end states of an anisotropic
Heisenberg model, and the discrepancies about the existence of spin 1/2
excitations without classical analog are removed.

We thank C. Saylor for the experimental data. K.H. and C.D.B. are supported
by CONICET, Argentina. A.A.A. is partially supported by CONICET. K. H.
thanks the Max-Planck Institute PKS (Dresden) for computational facilities.
This work was supported by PICT 03-00121-02153 of ANPCyT and PIP 4952/96 of
CONICET.


\end{document}